\newcommand{\lpx}{\stackrel{\leftarrow}{\partial}_{x}}
\newcommand{\rpx}{\stackrel{\rightarrow}{\partial}_{x}}
\newcommand{\lpp}{\stackrel{\leftarrow}{\partial}_{p}}
\newcommand{\rpp}{\stackrel{\rightarrow}{\partial}_{p}}
\newcommand{\ben}{\begin{equation}}
\newcommand{\een}{\end{equation}}
\newcommand{\bea}{\begin{eqnarray}}
\newcommand{\eea}{\end{eqnarray}}
\newcommand{\nn}{\nonumber\\ }
\newcommand{\dq}{deformation quantization}
\newcommand{\cW}{{\cal W}}
\newcommand{\vt}{{\vartheta}}
\renewcommand{\*}{\star}
\renewcommand{\sp}{$\star $-product}
\documentclass{article}

\usepackage{epsfig}
\begin{document}

\parskip=4pt
\baselineskip=14pt

%%%%%%%%%%%% Title %%%%%%%%%%

\title{Quantum Mechanics Another Way}
\author{J. Hancock, M. A. Walton\\\\{\it Department of Physics,
University of Lethbridge}\\
{\em Lethbridge, Alberta, Canada\ \  T1K 3M4}\\
{\small jason.hancock@uleth.ca, walton@uleth.ca}\\\\
and\\\\
B. Wynder\\\\{\it Department of Physics,
University of Alberta}\\
{\em Edmonton, Alberta, Canada\ \ T6G 2J1 }\\
{\small bwynder@phys.ualberta.ca}}

\maketitle
%%%%%%%%%%%%%%%%%%%%%%%%%%%%%%%%%%%%%%%%%%%%%%%%%%%%%%%%%%%%%%%%%%%%%%%%%%%%%%%%%%%%%%%%%%%%%%%%%%%%%%%%%%%%%%%%
\begin{abstract}
Deformation quantization (sometimes called phase-space
quantization) is a formulation of quantum mechanics that is not
usually taught to undergraduates. It is formally quite similar to
classical mechanics: ordinary functions on phase space take the
place of operators, but the functions are multiplied in an exotic
way, using the \sp. Here we attempt a brief, pedagogical
discussion of deformation quantization, that is suitable for inclusion
in an undergraduate course.

\end{abstract}
%%%%%%%%%%%%%%%%%%%%%%%%%%%%%%%%%%%%%%%%%%%%%%%%%%%%%%%%%%%%%%%%%%%%%%%%%%%%%%%%%%%%%%%%%%%%%%%%%%%%%%%%%%%%%%%%
\section{Introduction}

Another way of doing quantum mechanics grew from pioneering works
of Weyl, Wigner, Groenewold, Moyal, Baker, and others. For
reviews, see \cite{DQrev}. Its importance as an autonomous
formulation of quantum mechanics was appreciated in the 1970s; it
was also understood that it could be viewed as a deformation of
classical mechanics \cite{BFFLS}. The formulation was therefore
dubbed {\it deformation quantization}.

The pedagogical review article \cite{HH} advocated that \dq\ be
included in graduate studies, and it is treated in some graduate
textbooks (see \cite{Bal}, for example). We believe undergraduate
students would also benefit by learning something about \dq, i.e.
from an exposure to it. For that reason, we attempt here a brief,
pedagogical discussion of \dq, with upper-level undergraduates and
their instructors in mind.

To make our presentation as pedagogical as possible, we will
restrict to the case of one degree of freedom, i.e., to the case
of a single particle moving on the $x$-axis. We'll only treat pure
quantum states, since the generalization to mixed states is
straightforward. Lastly, although the Weyl ordering and
corresponding Groenewold-Moyal \sp\ determine just one of many
different ways to do deformation quantization, we will only
discuss the Weyl-Groenewold-Moyal case.

It is important to point out, however, that deformation
quantization is not just of pedagogical interest. Recently, it has
been an active research topic in both physics and mathematics.
Physicists studying string theory use the methods of deformation
quantization because in certain conditions, strings live in spaces
whose coordinates do not commute \cite{DH,Sch}, much as the two
quantum coordinates $X,P$ of phase-space do not, according to
(\ref{CXP}). The mathematician Kontsevich's work on deformation
quantization in \cite{Kon} was part of the reason he was awarded
the Fields medal, math's highest honor.

\section{Overview}

Consider a single particle, with position $x$ and momentum $p$.
Phase space is the two-dimensional space with coordinates $(x,p)$.
Each point of phase space specifies a classical state of the
system, and as a state evolves in time, a point traces out a path
in phase space. For example, a simple harmonic oscillator follows
an elliptical trajectory, centered on the origin $(x,p)=(0,0)$.

If our knowledge of a system were imprecise, the state might be
given as a probability distribution on phase space, perhaps a
bump with its center at the most probable values $(\bar x,\bar
p)$.  Classical dynamics could be done by following the evolution
of this distribution.

Can quantum mechanics be done in a similar way? The answer is yes,
and the way is called deformation quantization \cite{DQrev,BFFLS}.

By doing quantum mechanics this way, the introduction of abstract
quantum states, and their Hilbert spaces, can be avoided. States
are described instead by functions on phase space, as in classical
mechanics. As a consequence, the relation between quantum and
classical mechanics may be understood better.

Before sketching how deformation quantization is done, we need to
emphasize that quantization of a classical system is not a unique
procedure, no matter what formulation of quantum mechanics is
used. In the operator approach, $x$ and $p$ are replaced by the
corresponding operators, denoted $X$ and $P$. Suppose we need to
work with something like $x^2p$ in quantum mechanics, should we
consider $X^2P$? Or $XPX$, for example? The ambiguity can be
reduced by demanding that the operators constructed be Hermitian,
but that does not eliminate the choice completely.

This operator-ordering ambiguity is not a new problem, special to
\dq, since it is part of the usual operator approach to quantum
mechanics. A choice must be made, so let's use the so-called Weyl
ordering, and write \ben {\vt} (x^2p)\ =\
\frac{1}{3}\,(X^2P+XPX+PX^2)\ . \label{Wx2p} \een Weyl ordering
can be extended to functions on phase space by specifying how it
works on all monomials $x^mp^n$, and then applying it to the
Taylor expansions of functions. $\vt(x^mp^n)$ is simply the
average of all possible orderings of $m$ factors of $X$ and $n$
factors of $P$.

$\vt$ is known as the {\it Weyl map}, taking functions on phase
space to operators. We now have an operator ${\vt}(f)$ that
corresponds to a function $f=f(x,p)$ on phase space. We'll call
such an operator a {\it Weyl operator}. Multiplying two Weyl
operators gives another one, so their algebra closes, as it must.

Remarkably, one can prove a stronger statement. It is this result
that makes deformation quantization possible. Groenewold showed
that \ben {\vt}(f)\, \vt(g)\ =\ \vt(f\*g)\ .\label{hom}\een That
is, multiplying two Weyl operators is equivalent to
$\*$-multiplying the corresponding phase-space functions, and then
applying $\vt$. The \sp\ (pronounced star-product) takes the form
\ben f(x,p)\*g(x,p)\ \equiv \ f(x,p)\,
\exp\Big\{\frac{i\hbar}{2}\, \big(\lpx\rpp-\lpp\rpx\big)\Big\}\,\,
g(x,p)\ .\label{Moy}\een Here $\partial_x \equiv
\frac{\partial}{\partial x}$, etc., and the arrows indicate the
directions in which the derivatives act. The exponential is to be
understood using the series expansion \ben e^A\ =\
\sum_{n=0}^\infty\,\frac{A^n}{n!}\ .\label{eexp} \een Since it is
directly related to the product of Weyl operators, the \sp\ is
non-commutative and associative, as the product of operators is.
It is a strange-looking product precisely because it must mimic
the product of operators.

Eqn.\ (\ref{hom}) is important because it suggests that one might
be able to avoid constructing operators from functions on phase
space and just work with the functions directly, as long as they
are multiplied using the \sp. This is exactly what is done in
deformation quantization. Operator products are changed to
$\*$-products, and the Weyl map $\vt$ is factored off, roughly
speaking.

To give the flavor of how it goes, it is easy to show from
(\ref{Moy}) that \ben x\* p\ =\ xp\ +\ \frac{i\hbar}{2}
\label{xsp}\een and \ben p\* x\ =\ xp\ -\ \frac{i\hbar}{2}\ .
\label{psx}\een Then the {\it $\*$-commutator}, or Moyal bracket,
of $x$ and $p$ is \ben [x,p]_\*\ \equiv \ x\* p-p\* x\ =\ i\hbar\
.\label{xscp}\een This result is consistent with the crucial
canonical commutation relation \ben [X,P]\ =\ i\hbar\
.\label{CXP}\een

To do quantum mechanics we need further ingredients, beyond the
$\*$-product. The quantum state must be described. In the operator
formulation, a {\it pure} quantum state is describable by a state
vector $|\psi\rangle$. The most general type of quantum state is a
{\it mixed} state, however, and it incorporates classical
probabilities for different pure states. It is a mixed quantum
state that corresponds to the classical distribution on phase
space mentioned in the second paragraph. To specify such a mixed
state, the density matrix (sometimes called the state operator)
$\hat\rho$ must be used (see \cite{CDL} for a nice discussion).

We'll nevertheless restrict this discussion to the case
of a pure state $\psi$,
since it makes the presentation easier to follow. Then \ben
\hat\rho\ =\ |\psi\rangle\, \langle\psi|\ \ . \label{rhopp} \een
Generalization to mixed states is straightforward.

Now we have an operator, and to do deformation quantization, we
need a function on phase space that corresponds to it in the
manner of (\ref{hom}). That is, we need an operation $\cW$ acting
on operators and giving phase-space functions, that satisfies \ben
\cW\big(\, \vt(f)\,\big)\ =\ f\ ,\ \ \label{CWf} \een for any
$f=f(x,p)$; we need $\cW=\vt^{-1}$. What works is \ben \cW\big(\,
G\,\big)\ =\ \hbar\, \int dy\, e^{-ipy}\, \langle x+\frac{\hbar
y}{2}|\,  G\, |x-\frac{\hbar y}{2}\rangle\ .\label{CTh} \een As
with $X,P$ vs. $x,p$, we will denote operators with upper-case
symbols, to distinguish them from phase-space functions. $\cW\big(
G\big)$ is called the {\it Weyl transform} (and sometimes the Weyl
symbol) of the operator $ G$.

The Weyl transform of the density matrix $\hat\rho=|\psi\rangle\,
\langle\psi|$, \ben \cW\big( \hat\rho\, \big)\ =\ \hbar\, \int
dy\, e^{-ipy}\, \psi^*(x-\frac{\hbar y}{2})\,\psi(x+\frac{\hbar
y}{2})\ , \label{Wig} \een is the central object in deformation
quantization. After normalization, it is known as the {\it Wigner
function}: \ben \rho\ \equiv\ \frac{\cW(\hat\rho)}{2 \pi \hbar}\ \
. \een It describes the quantum state of the system, and all
observable probabilities can be calculated from it.

This hints at a punchline: deformation quantization is the Weyl
transform of quantum mechanics done with the density matrix, or
state operator.

%%%%%%%%%%%%%%%%%%%%%%%%%%%%%%%%%%%%%%%%%%%%%%%%%%%%%%%%%%%%%%%%%%%%%%%%%%%%%%%%%%%%%%%%%%%%%%%%%%%%%%%%%%%%%%%%
\section{Weyl Transform and Groenewold-Moyal Star Product}

In this section, some detail and proofs omitted in the previous
section will be provided. It can be skipped in a first reading.

First, consider the Weyl map of functions on phase space, like
$f(x,p)$. By expanding, it is not hard to convince oneself that
\ben \vt\left((ax+bp)^n\right)\ =\ (aX+bP)^n\ \ , \label{Wmon}\een
for all parameters $a,b$. It follows that \ben {\vt}(e^{ax+bp})\
=\ e^{aX+bP}\ \ ,\label{Wee} \een using (\ref{eexp}). Now, the
Taylor series of $f(x,p)$ about $(x,p)=(0,0)$ can be written as
\ben f(\partial_a,\partial_b)\, e^{ax+bp}\,|_{a,b=0}\ =\ f(x,p)\
.\label{Tay}\een This combined with (\ref{Wee}) gives \ben \vt(f)\
=\ f(\partial_a,\partial_b)\, e^{aX+bP}|_{a,b=0}\ \
,\label{Wfe}\een a useful general formula for Weyl operators.
Slightly different versions of the same formula are \ben \vt(f)\
=\ f(a,b)\, e^{{\stackrel{\leftarrow}{\partial}_{a}}X+
{\stackrel{\leftarrow}{\partial}_{b}}P} \,|_{a,b=0}\ =\
e^{{\stackrel{\rightarrow}{\partial}_{a}}X+
{\stackrel{\rightarrow}{\partial}_{b}}P}\, f(a,b)\,|_{a,b=0}\ .
\label{Wfev}\een

Another formula that one sees quite often is \ben {\vt}(f)\ =\
\frac{1}{(2 \pi)^2} \int d\tau \, d\sigma \, dx\, dp \ f(x,p)
e^{i\tau(P-p) + i\sigma(X-x)}\ . \label{Wfint}\een It follows from
(\ref{Wee}) and Fourier methods. The usual Fourier expression is
\ben f(x,p)\ =\
 \int d\tau \, d\sigma \ \tilde f(\tau,\sigma)
e^{i\sigma x + i\tau p}\ \ , \label{fxp}\een where \ben\tilde
f(\tau,\sigma)\ =\ \frac{1}{(2 \pi)^2} \int dx' \, dp' \, \
f(x',p') e^{-i\tau p' - i\sigma x'} \label{tfF}\een is the Fourier
transform of $f(x,p)$. According to (\ref{Wee}), the Weyl map
$\vt$ simply replaces $e^{i\sigma x + i\tau p}$ with $e^{i\sigma X
+ i\tau P}$. Making that replacement in (\ref{fxp},\,\ref{tfF})
gives (\ref{Wfint}).

Equations (\ref{Wfe},\,\ref{Wfev},\,\ref{Wfint}) are not so useful
for calculating simple examples. They are, however, important for
discussing the general properties of Weyl maps. For example, the
\sp\ is introduced because of the crucial property (\ref{hom}). In
order to prove it, (\ref{Wfev}) can be used. First, \ben
\vt(f)\vt(g)\ =\
 f(a,b)\,e^{{\stackrel{\leftarrow}{\partial}_{a}}X+
{\stackrel{\leftarrow}{\partial}_{b}}P}\,
e^{{\stackrel{\rightarrow}{\partial}_{\bar a}}X+
{\stackrel{\rightarrow}{\partial}_{\bar b}}P}\,g(\bar a,\bar b)  \
|_{a,b,\bar a,\bar b=0}\ .\label{cWfgi}\een Now use the
(simplified) Baker-Campbell-Hausdorff formula, \ben e^A\, e^B\ =\
e^{\frac{1}{2}[A,B]}\,e^{A+B}\ ,\label{BCH}\een valid when the
commutator $[A,B]$ commutes with both $A$ and $B$. Eqn.
(\ref{cWfgi}) becomes \ben \vt(f)\vt(g)\ =\
 e^{(\partial_a+\partial_{\bar a})X+(\partial_b+\partial_{\bar b})P}\,f(a,b)\,
 e^{\frac{i\hbar}{2}\,({\stackrel{\leftarrow}{\partial}_{a}}
 {\stackrel{\rightarrow}{\partial}_{\bar b}}-
 {\stackrel{\leftarrow}{\partial}_{b}}
 {\stackrel{\rightarrow}{\partial}_{\bar a}})}\, g(\bar
a,\bar b)  \ |_{a,b,\bar a,\bar b=0}\ , \label{cWfgii}\een or \ben
\vt(f)\vt(g)\ =\
 e^{\partial_aX+\partial_bP}\,f(a,b)\,
 e^{\frac{i\hbar}{2}\,({\stackrel{\leftarrow}{\partial}_{a}}
 {\stackrel{\rightarrow}{\partial}_{ b}}-
 {\stackrel{\leftarrow}{\partial}_{b}}
 {\stackrel{\rightarrow}{\partial}_{a}})}\, g(a,b) \ |_{a,b=0}\ .
 \label{cWfgiii}\een The result
(\ref{hom}) then follows.

As mentioned above, the Groenewold-Moyal \sp, defined in
(\ref{Moy}), is non-commutative, i.e. $f\*g\not=g\*f$. It is,
however, associative \ben (f_1\*f_2)\*f_3\ =\ f_1\*(f_2\*f_3)\
.\label{asso}\een It shares those properties with the product of
operators, as (\ref{hom}) demands. Similarly, it is easy to show
that \ben \left(f_1\*f_2\right)^*\ =\ f_2^*\*f_1^*\
,\label{adj}\een in agreement with the rule $(F_1F_2)^\dagger\,=\,
F_2^\dagger F_1^\dagger$ for operators $F_1,\, F_2$.

The exponent of the \sp\ (\ref{Moy}) indicates the most important
property of \dq: its intimate relation to classical physics. In
classical mechanics, it is the Poisson bracket of functions on
phase space, \ben \{f,g\}\ =\ \frac{\partial f}{\partial x}\,
\frac{\partial g}{\partial p}\ -\ \frac{\partial f}{\partial p}\,
\frac{\partial g}{\partial x}\ =\ f\,(\lpx\rpp-\lpp\rpx)\,g\
\label{Poi}\een that enters the dynamical equations \cite{Gol}. In
the operator formulation of quantum mechanics, it is the
commutator $[F,G]$ of operator observables $F$ and $G$ that is
important. In \dq, the $\*$-commutator \ben  [f,g]_\*\ \equiv \ f\* g -
g\* f \label{FscG}\een (recall eqn. (\ref{xscp})) of functions $f$
and $g$ takes its place. The equation \ben \lim_{\hbar \rightarrow
0}\frac{1}{i\hbar}[f,g]_{\* } = \{ f,g \}\ \ \label{hlim}\een
encodes the relation between classical and quantum mechanics in
\dq.

Now let us consider the Weyl transform $\cW( G)$ of an operator $ G$,
defined by the property (\ref{CWf}), and given explicitly by the
formula (\ref{CTh}). Of course, here $ G =  G(X,P)$ is an operator
function of the operators $X$ and $P$.

First, let us argue for the formula (\ref{CTh}). From the
development preceding (\ref{Wfe}), it is clear that we only need
to show \ben \cW(e^{aX+bP})\ =\ e^{ax+bp}\ , \label{cee}\een i.e.,
the inverse of (\ref{Wee}). This is not difficult, however, using
the Baker-Campbell-Hausdorff formula (\ref{BCH}), and that
$e^{bP}$ is a translation operator, i.e., $e^{bP}|x\rangle =
|x+i\hbar b\rangle$.

Finally, we should note that there is an
inverse analog of (\ref{hom}): \ben \cW( G)\*\cW( L)\ =\ \cW( G L)\ .
\label{cscc}\een Because of (\ref{CWf}), it works if the operators
are Weyl operators, i.e., if $ G=\vt( g)$ and $ L=\vt(\ell)$ for
some phase-space functions $ g$ and  $\ell $.  They will be, however:
\ben X^2 P^2 X\ =\
\vt(x)\vt(x)\vt(p)\vt(p)\vt(x)\ =\ \vt(x\*x\*p\*p\*x)\ , \een for
example.

%%%%%%%%%%%%%%%%%%%%%%%%%%%%%%%%%%%%%%%%%%%%%%%%%%%%%%%%%%%%%%%%%%%%%%%%%

\section{Wigner Function}

Now that the basics have been established, we can consider the
description of quantum states and their evolution. As stated
above, the central object is the Wigner function (\ref{Wig}).

In the operator method, one must determine the description of a
state. Its state vector $|\psi\rangle$, the description, is found
by solving the Schr\"odinger equation
$\frac{d}{dt}|{\psi}\rangle=|\stackrel{.}{\psi}\rangle=\frac{1}{i\hbar}
\hat{H}|\psi\rangle$. In a similar way, the starting point in
deformation quantization is the dynamical equation for the Wigner
function $\rho$. From (\ref{Wig}), we find
\begin{eqnarray}
\frac{\partial \rho}{\partial t}\ =\ \frac{1}{2 \pi} \int
\!\!\!\!&dy&\!\!\!\!\, e^{-iyp}\, \Big\{\,\,\!\! \langle
x+\frac{\hbar
y}{2}|\stackrel{.}{\psi}\rangle\langle\psi|x-\frac{\hbar
y}{2}\rangle \nonumber
\\ &&\ +\ \langle x+\frac{\hbar y}{2}|\psi\rangle\langle\stackrel{.}{\psi}|x-\frac{\hbar
y}{2}\rangle\!\!\!\,\,\Big\}\ \ . \label{dynshro}
\end{eqnarray}
Substituting the Schr\"odinger equation (and its adjoint) then
gives
\begin{eqnarray}
\frac{\partial \rho}{\partial t}\ &=&\ \frac{1}{i\hbar}\,\,
\cW\left([\hat{H},\hat{\rho}]\right)\nonumber\\
                &=&\ \frac{1}{i\hbar}\,
                [H,\rho]_\*\ .
\end{eqnarray}
In the last step, we used (\ref{cscc}), and we have defined the
Weyl transform $H \equiv \cW(\hat H)$ of the operator Hamiltonian
$\hat H$.

For stationary states, $\frac{\partial\rho}{\partial t}=0$, so
that \ben [H,\rho]_{\*}\ =\  0\ . \label{Hrcom}\een Thus the
Hamiltonian and Wigner function $\*$-commute. A stronger relation
can be derived more directly. The Schr\"odinger equation
simplifies for stationary states to $\hat
H|\psi\rangle=E|\psi\rangle$, where $E$ is the energy. This
implies that $\hat H\hat\rho = E\hat\rho$, which Weyl transforms
to
\begin{equation} H \* \rho = E\rho\ \ . \label{HrEr}\end{equation}
In the next section, this simplified dynamical equation will allow
us to solve for the Wigner function of the stationary states of
the simple harmonic oscillator.

Once the Wigner function is determined, how is it used? First of
all, by (\ref{Wig}), the probability densities are \bea {\cal
P}(x)\ &=&\ |\psi(x)|^2\ =\ \int dp\ \rho(x,p)\ , \nn {\cal P}(p)\
&=&\ |\psi(p)|^2\ =\ \int dx\ \rho(x,p)\ . \label{xpden}\eea
Clearly then, the Wigner function is normalized and real: $\int
dx\, dp\, \rho = 1$ and $\rho^{\ast}=\rho$.

All observable expectation values can be calculated using the
Wigner function. The expectation value of an operator $G$ is \ben
\langle G\rangle\ =\ \langle\psi|G |\psi\rangle\ =\ {\rm
Tr\,}(\hat\rho\, G)\ .\label{eTr}\een In deformation quantization,
this translates into \ben  \langle G\rangle\ =\ \int dx\,dp\,\,
\rho\* g\ \ , \label{eGints}\een where $g \equiv \cW(G)$. Roughly,
one can think of the integral over phase space as the analog of
the trace, and as discussed above, the star product takes the
place of the operator product. The important cyclic property of a
trace is encoded in \ben \int dx\,dp\,\, f\* g\ =\ \int dx\,dp\,\,
g\* f\ =\ \int dx\,dp\,\, f\,g\ .\label{cTr}\een

%%%%%%%%%%%%%%%%%%%%%%%%%%%%%%%%%%%%%%%%%%%%%%%%%%%%%%%%%%%%%%%%

\section{Example: Simple Harmonic Oscillator}

The most common non-trivial example studied in physics, quantum or
classical, is the simple harmonic oscillator (SHO). We will now
treat the quantum SHO using deformation quantization, following
\cite{CFZ}. Our results will illustrate some of the properties of
the Wigner function in deformation quantization.

Recall from above that $H\* \rho=E\rho$. Taking $\omega = m =1$
for simplicity, the SHO Hamiltonian is $\hat
H=\frac{1}{2}(P^2+X^2)$. To use (\ref{HrEr}), we need to calculate
\begin{equation}
H\*\rho\ =\ \sum_{n=0}^{\infty}(\frac{i\hbar}{2})^{^{{\scriptstyle
n}}} \frac{1}{n!}\, H\,(\lpx \rpp-\lpp\rpx)^n\,\rho \label{spsum}
\end{equation}
with $H=\cW(\hat H)= \frac{1}{2}(p^2+x^2)$. Since $H$ is quadratic
in both $x$ and $p$, the sum in (\ref{spsum}) only has to range
from 0 to 2. $H\*\rho= E\rho$ therefore yields \ben \Big[ x^2+p^2
- \frac{\hbar^2}{4}(\partial_p^2 +
\partial_x^2)+i\hbar(x\partial_p-p \partial_x)-2E\Big]\,\rho\ =\
0\ .\label{endsho}
\end{equation}
Separating (\ref{endsho}) into its real and imaginary parts
reveals two partial differential equations,
\begin{eqnarray}
(x\partial_p - p\, \partial_x)\rho\  =\ 0  \label{imag}
\end{eqnarray}
and
\begin{eqnarray}
\Big[ x^2+p^2 -\frac{\hbar^2}{4}(\partial_p^2 +
\partial_x^2)-2E\Big]\,\rho\ =\ 0 \ \ .\label{real}
\end{eqnarray}
(\ref{imag}) shows that $\rho(x,p)$ is a function only of
$x^2+p^2$. It'll be more convenient to use
$u={2(x^2+p^2)}/{\hbar}=4H/{\hbar}$. We write $\rho(x,p)=\rho(u)$,
and substitute into (\ref{real}). The chain rule gives
\begin{equation}
\partial_x^2 \ = \ \frac{\partial}{\partial x}
\left(\frac{\partial u}{\partial x}\frac{\partial}{\partial
u}\right) \ =\
\frac{4}{\hbar}\partial_u+\frac{16x^2}{\hbar^2}\partial_u^2\ ,
\label{diex2}
\end{equation}
and \ben
\partial_p^2\ =\ \frac{4}{\hbar}\partial_u+\frac{16p^2}{\hbar^2}
\partial_u^2 \label{diep2}
\een similarly. Substituting into (\ref{real}) gives \ben 0 \ = \
\bigg(\frac{u}{4}-\partial_u-u\,
\partial_u^2-\frac{E}{\hbar}\bigg)\rho(u)\ \ .\label{realend}
\een This is the differential equation to be solved to determine
the Wigner function. It is no more difficult to solve than the
Schrodinger equation in the operator formulation (but does not
lead to Hermite polynomials, as we'll see).

Setting $\rho(u)=e^{-\frac{u}{2}}L(u)$ simplifies considerations.
Substituting into (\ref{realend}) gives \ben 0 \ = \ \Big[\, u\,
\partial_u^2 +(1-u)\partial_u+ \bigg(\frac{E}{\hbar}-\frac{1}{2}\bigg)
\,\Big]\,L(u)\ . \label{Lageq}\een We can look for a series solution, by
substituting $L(u) = \sum_{a=0}^\infty\, \ell_a u^a$. The resulting
recursion relation is \ben (a+1)^2\ell_{a+1}\ =\
\bigg(a-\frac{E}{\hbar}+\frac{1}{2}\bigg)\ell_a\ .\label{recur}\een
For a normalizable solution, we need $L(u)$ to be a polynomial. That is,
$\ell_a$ must vanish for all $a$ greater than some finite $n$. The recursion
relation tells us this will happen if $E = (n+\frac{1}{2})\hbar$, for some
non-negative integer $n$. These are exactly the SHO quantum
energies (recall that
we set $\omega=1$).

The recursion relation yields solutions $L_0=1$, $L_1=1-u$,
$L_2=1-2u+\frac{1}{2}u^2$, etc. The normalization constant can be
fixed by requiring $\int dx\,dp\,\rho = 1$. The final general
result is
\begin{eqnarray}
\rho_n\ =\ \frac{(-1)^n}{\pi} e^{-\frac{2H}{\hbar}}
L_n({4H}/{\hbar})\ ,
\end{eqnarray}
where $L_n$ denotes the $n$th Laguerre polynomial.

The SHO can also be solved in algebraic fashion, as in operator
quantum mechanics. Defining \ben a\ \equiv\
\frac{1}{\sqrt{2\hbar}}\,(x+ip)\, ,\ \ \ \ \ \ \ a^*\ \equiv\
\frac{1}{\sqrt{2\hbar}}\,(x-ip)\ \ ,\label{ladd}\een we can write
$H=\hbar\left( a^*\* a + \frac{1}{2}\right)$. The ladder functions
satisfy \ben [a,a^*]_\*\ =\ 1\ ,\ \ \ \ [H,a^*]_\*\ =\ \hbar\
.\een The SHO ground state is described by a Wigner function
obeying \ben a\*\rho_0\ =\ \rho_0\* a^*\ =\ 0\ .\een The form of
the zeroth Wigner function can be found directly from these
equations. The others are found using \ben \rho_n\ =\ a^{\*
n}\*\rho_0\*(a^*)^{\* n}\ \ .\een

%\pagebreak
\begin{figure}[h]
\begin{center}
\epsfig{file=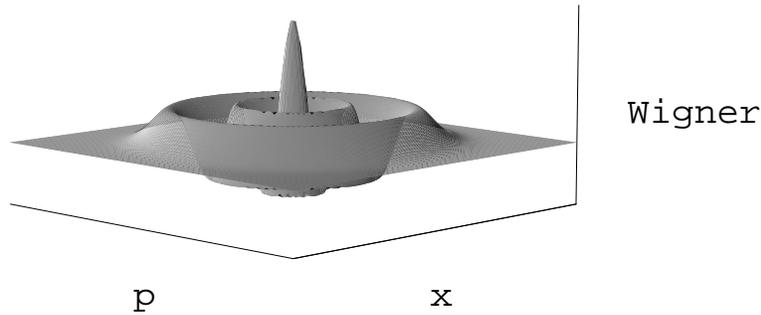,width=4in,angle=0} \caption{SHO
Wigner Function for $n=4$} \label{plot:wigner} \hspace*{0.5in}
\end{center}
\end{figure}

\begin{figure}[h]
\begin{minipage}{2.3in}
\epsfig{file=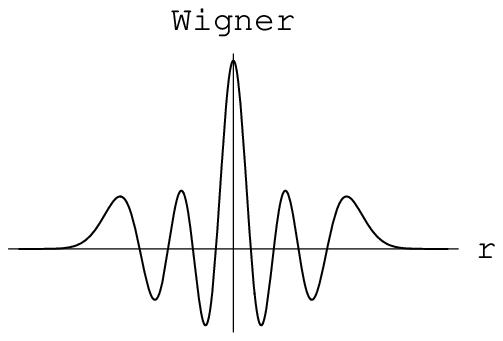,width=2.3in,angle=0} \caption{Wigner
function for $n=4$, $r^2=x^2+p^2$} \label{r2}
\end{minipage}
\hspace*{0.5in}
\begin{minipage}{2.5in}
\epsfig{file=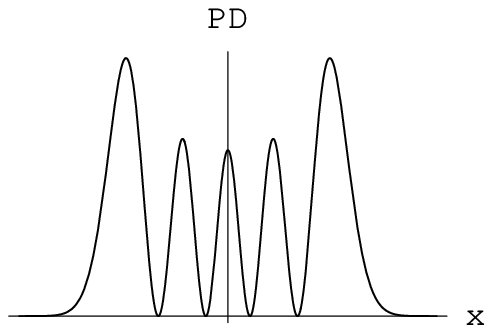,width=2.5in,angle=0}
\caption{Probability density for $x$ for SHO, $n=4$}
\label{density}
\end{minipage}
\end{figure}

Figures \ref{plot:wigner} and \ref{r2} depict the Wigner function
for the $n=4$ stationary state of the SHO. Because it only depends
on $u$, it has the circular symmetry seen in Figure
\ref{plot:wigner}. Figure \ref{r2} therefore gives the Wigner
profile for any straight line in phase space passing through the
origin.

The classical SHO with this symmetry follows a circular trajectory
in phase space, centered on the origin. The naive expectation might
be then that the Wigner function, describing the corresponding
quantum state, is a spread-out version of this; a single, circular
ridge located above the corresponding classical phase path. The
Figures show that only part of the Wigner function looks like
that. There are oscillations in $\rho$, and it even goes into
negative values. This means it is not a true probability
distribution; it is instead called a quasi-probability
distribution.

This feature partly explains why \dq\ is less popular than other
formalisms of quantum mechanics. We stress again, however, that it
reproduces all the predictions of the more familiar operator
methods. For instance, Figure \ref{density} plots the probability
density ${\cal P}(x)$ against $x$ (it could also be ${\cal P}(x)$
against $p$, because of symmetry), found using (\ref{xpden}). The
curve is identical to that found from the state vector solving
Schrodinger's equation. It is also true that expectation values of
operators, such as $X^mP^n$, calculated using (\ref{eGints}),
agree perfectly with those calculated in the more familiar
formulations of quantum mechanics.

%%%%%%%%%%%%%%%%%%%%%%%%%%%%%%%%%%%%%%%%%%%%%%%%%%%%%%%%%%%%%%%%%%%%%%%%%%%%%%%%%%%%%%%%%%%%%%%%%%%%%%%%%%%%%%%
\section{Conclusion}

Deformation quantization is another way to do quantum mechanics.
It has some strange features, such as a quasi-probability
distribution, and an exotic way of multiplying functions, the \sp.
But it is perfectly consistent, and its predictions agree with
those made using other formulations of quantum mechanics.
Furthermore, it is independent of the other formulations. On another
planet, it might
be the method of doing quantum mechanics discovered first! \cite{Zl}

We believe that studying it, along with other quantum methods,
deepens understanding of quantum physics. For example, writing
$FG$ might allow one to slip into thinking that the product of
operators is similar to an ordinary product of functions. But the
\sp\ simulates the product of operators, in the way discussed
above, and so eqn. (\ref{Moy}) makes clear that products of
operators are tricky things.

Deformation quantization is a bit of a well-kept secret,
especially at the undergraduate level. We hope our brief
discussion of it here can serve as a gentle introduction to the
subject for upper-level undergraduates, and perhaps others.
\vskip.5cm
\noindent{\bf Acknowledgments}\hfill\break
This work was completed while JH and BW were undergraduate summer research assistants
at the University of Lethbridge.
For funding we thank the University of Lethbridge
Research Fund and NSERC of Canada.

%%%%%%%%%%%%%%%%%%%%%%%%%%%%%%%%%%%%%%%%%%%%%%%%%%%%%%%%%%%%%%%%%%%%%%%%%%%%%%%%%%%%%%%%%%%%


\begin{thebibliography}{99}

\bibitem{DQrev} D.B. Fairlie, ``The formulation of quantum mechanics in terms of
phase space functions,'' Proc. Cambridge Phil. Soc. {\bf 60},
581-586 (1964),\\ M.V. Berry, ``Semi-classical mechanics in phase
space: a study of Wigner's function,'' Philos. Trans. Roy. Soc.
London Ser. A {\bf 287}, 237-271 (1977), \\ N.L. Balazs, B.K.
Jennings, ``Wigner's function and other distribution functions in
mock phase spaces,'' Phys. Rept. {\bf 104}, 347--391 (1984), \\ M.
Hillery, R. O'Connell, M. Scully, E. Wigner, `` Distribution
functions in physics: fundamentals,''
Phys. Rept. {\bf 106}, 121--167 (1984),\\
H.-W. Lee, ``Theory and applications of the quantum phase-space
distribution functions,'' Phys. Rept. {\bf 259}, 147-211 (1995), \\
C. Zachos, ``Deformation quantization: quantum mechanics lives and
works in phase-space,'' Int. J. Mod. Phys. {\bf A17} (3), 297-316
(2002) [hep-th/0110114]
\bibitem{BFFLS} F. Bayen, M. Flato, C. Fronsdal, A. Lichnerowicz,
D. Sternheimer, ``Deformation theory and quantization I, II,''
Ann. Phys. (N.Y.) {\bf 111}, 61, 111 (1978)
\bibitem{HH}A. Hirshfeld, P. Henselder, ``Deformation quantization in the teaching of
quantum mechanics,"  Am. J. Phys.  {\bf 70}, 537 (2002)
[quant-ph/0208163]
\bibitem{Bal} L. Ballentine, {\em Quantum mechanics: a modern
development} (World Scientific, 1998), chapter 15
%%%%%%%%%%%%%%%%
\bibitem{DH} M. R. Douglas, C. Hull, ``D-branes and the noncommutative
torus," J. High Energy Phys.  {\bf 9802}, 008 (1998)
[hep-th/9711165]
\bibitem{Sch} V. Schomerus, ``D-branes and deformation quantization,"
J. High Energy Phys.  {\bf 9906},  030 (1999) [hep-th/9903205]
\bibitem{Kon} M. Kontsevich, ``Deformation quantization of Poisson manifolds,"
q-alg/9709040 (1997)
\bibitem{CDL} C. Cohen-Tannoudji, B. Diu, F. Lalo\"e, {\em Quantum
mechanics} (Wiley, 1977), vol. I, complement E$_{\rm III}$
\bibitem{Gol} H. Goldstein, {\em Classical Mechanics}
(Addison-Wesley, 1980), 2nd ed., sections 9-4 \& 9-5
\bibitem{CFZ} T. Curtright, D. Fairlie, C. Zachos, ``Features of time-independent Wigner
functions,"  Phys. Rev.  {\bf D58} (2),  025002-025017 (1998)
[hep-th/9711183]
\bibitem{Zl} C. Zachos, ``Deformation Quantization: Quantum Mechanics Lives \& Works in
Phase-Space," Fermilab Colloquium Lecture, August 1, 2001;
streaming video available from www.fnal.gov/faw/seminars.html


\end{thebibliography}
\end{document}